# Rethinking Battery-free Sensing Communication via Wake-up Radios

Gaosheng Liu

*Abstract*—Battery-free sensors expose short, stochastic communication windows and may lose timing state whenever their main energy domain browns out. MagPie combines a microampere wake-up radio (WuR) with a separately backed low-power real-time clock (LP-RTC): the WuR widens the first-contact window, while the LP-RTC preserves the acquired phase for later exchanges. Energy gates, epoch-versioned schedules, and idempotent slot allocation extend this mechanism to static, single-hop All-to-One collection without a powered control anchor. The analysis accounts for role selection and duty-cycled listening, and bounds scheduled-retry tails only under explicit conditional quantile coverage. Evaluation separates three scopes. In independent, administratively censored simulation trials, MagPie completes 100/100 first rendezvous events in each of five trace-parameterized harvesting scenarios; Find completes 36–100/100. A single-collision-domain slotted-Aloha study shows that adaptive $K$ is necessary at high contender density and reports mean, P95, and confidence intervals through 120 components. Finally, controlled STM32WL33 experiments validate alignment, clock persistence, and six-device slot execution, including an 11.05-hour functional run. The study does not claim measured end-to-end energy, ambient-harvesting performance, or multi-hop scalability; highly variable harvesting still limits collection because coordination cannot create missing energy.

## I. Introduction

BATTERY-FREE sensors harvest ambient radio-frequency, solar, or kinetic energy instead of carrying a battery. They can reduce maintenance and electronic waste in pervasive sensing deployments [1], [2], and are an important part of the emerging Ambient IoT vision [3]. Their communication problem, however, differs fundamentally from conventional duty cycling. A small capacitor powers the device only briefly; after voltage falls below a shutdown threshold, the processor and main radio lose power until the harvester replenishes the buffer.

Figure 1 illustrates the resulting temporal asymmetry. Active windows can be only a few milliseconds, whereas charging intervals are two to five orders of magnitude longer [4], [5]. Intermittent-computing systems preserve local progress through checkpointing or idempotent tasks [6]–[10], but a checkpoint cannot make two independently powered radios awake at the same time. Communication therefore remains a system-level bottleneck.

Gaosheng Liu is with the Department of Information Engineering and Computer Science, University of Trento (e-mail: lgselite@163.com).

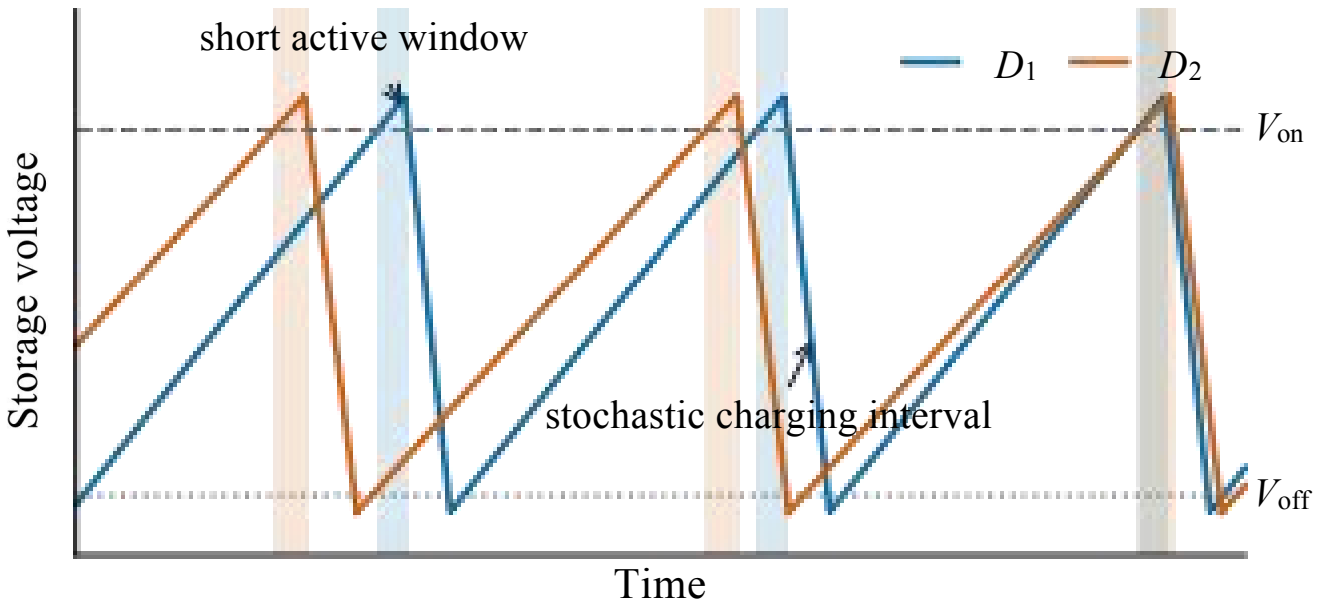

Fig. 1. Intermittent operation. A millisecond-scale active window is separated from the next by a much longer, stochastic charging interval. Main-radio communication requires two active windows to overlap.

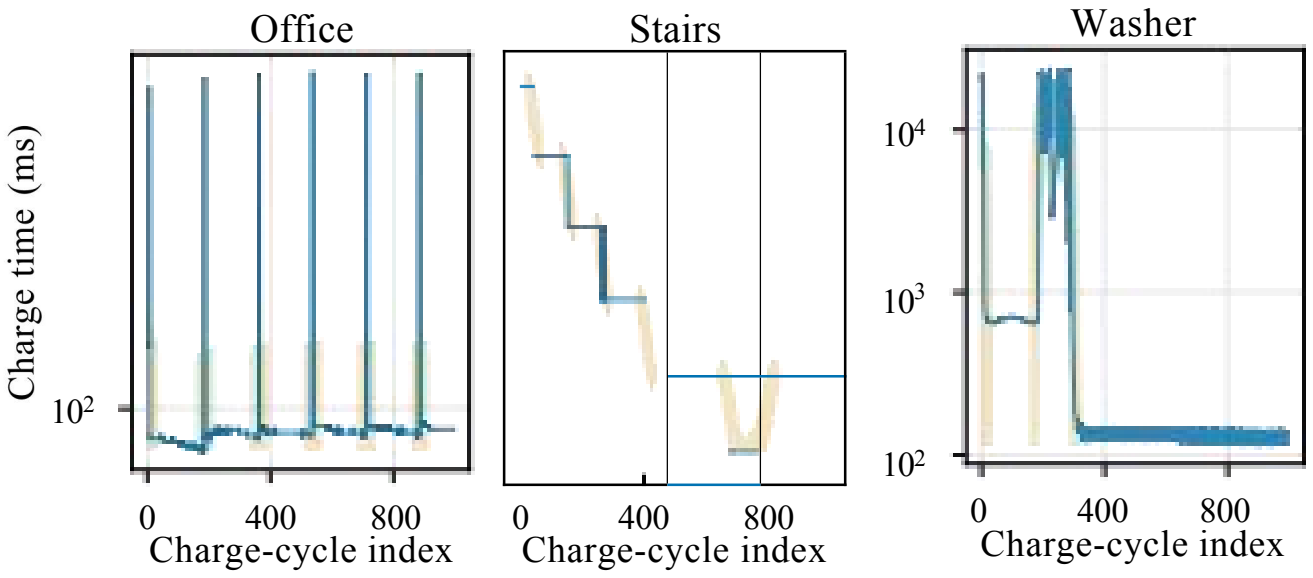

Fig. 2. Charging-time traces from the Bonito dataset [11]. The vertical axis is logarithmic; the orange band is a 60-cycle rolling mean ± one standard deviation. Variability differs markedly across source and location.

The difficulty is not merely a long mean charging time. Figure 2 shows that charging varies across cycles and devices even within one harvesting setting. Find and Bonito predict or learn future charging behavior [4], [5]; a prediction error can invalidate the intended rendezvous. Flync exploits mains-frequency light flicker, but only where that shared periodic signal exists [4]. A second family instead imposes a slotted schedule from a powered beacon or coordinator [12]–[14]. This can be fast, but reintroduces powered infrastructure and a distinguished point of availability.

These approaches expose two requirements. First, an infrastructure-free pair must discover one another despite stochastic, non-overlapping active windows and then retain the acquired timing state. Second, a network must turn pairwise contacts into a recoverable collection schedule without assuming that every device is powered in every round.

**Key idea.** Instead of predicting a narrow main-radio window, MagPie uses a WuR to widen it. Let $X > 0$ be a receiver's charging interval, $w$ its main-radio active width, $\delta$ its WuR listening duty factor, and $K$ the identifier-hashed role modulus. Conditional on $X$, assume the peer's beacon phase is uniform over that interval. Exactly one endpoint of a pair transmits

TABLE I
COORDINATION MECHANISMS AND BROWNOUT SEMANTICS. "EXTERNAL" MEANS THAT AN ALWAYS-ON ANCHOR RECREATES TIMING; IT IS NOT LOCAL PERSISTENCE.

| Protocol | Rendezvous basis | Powered anchor | After node brownout | Scope |
|---|---|---|---|---|
| Find [4] | random prediction | No | re-learn | pair |
| Flync [4] | flicker lock | No | re-lock | pair |
| Bonito [5] | learned prediction | No | re-learn | pair |
| CATO [15] | addressed WuR | No | rediscover | ND/MAC |
| Swift [12] | slotted beacon | Yes | external | network |
| Pulsar [13] | slotted beacon | Yes | external | collection |
| FreeBeacon [14] | slotted beacon | Yes | external | collection |
| MagPie (ours) | WuR + LP-RTC | No | local restore | collection |

with probability $p_{role} = 2K^{-1}(1 - K^{-1})$, giving

$$P_{hear} = p_{role}\,\delta\,\mathbb{E}\left[\min\left\{1, \frac{T_{listen} + w}{X}\right\}\right] \approx p_{role}\,\delta\,(T_{listen} + w)\,\mathbb{E}[X^{-1}], \quad (1)$$

where the second line applies only when $(T_{listen} + w)/X < 1$ with high probability. The expectation is over the interval distribution; replacing it with $1/\mathbb{E}[X]$ is generally incorrect. Equation (1) is not distribution free: correlated or nearly periodic charging can violate the uniform-phase premise and create a beat frequency with the LP-RTC tick. The role hash breaks sender symmetry but not such phase locking; repeated misses must therefore trigger phase re-estimation (and, in a deployed implementation, an identifier-derived tick dither). Listening is sustainable only while the harvester covers WuR and leakage current; otherwise $\delta < 1$ trades reception probability for charge progress.

MagPie builds on this observation. A WuR handshake establishes relative phase, and an LP-RTC in a separately backed domain preserves it while the MCU and main radio are off. Subsequent energy misses do not erase timing state. Multi-threshold voltage gates prevent a radio phase from starting below its configured budget, and a sender-owned epoch reconciles cycle updates after missed notifications. At network scale, component representatives contend to construct a versioned merge tree; a two-pass range allocation then assigns collision-free collection slots without a battery-powered control anchor.

The novelty is not the WuR primitive by itself, nor a new general-purpose routing metric. It is the hardware–protocol co-design that (i) listens during charging, (ii) separates timing persistence from the main energy domain, (iii) admits work only when a characterized voltage interval can fund it, and (iv) turns one probabilistic contact into a recoverable schedule. We deliberately scope the network layer to stationary, single-hop All-to-One collection. Multi-hop routing and mobility are not evaluated and are stated as limitations rather than implied capabilities.

Our contributions are as follows:

- We design a WuR–LP-RTC communication stack for intermittently powered devices, including energy gating, phase acquisition, persistent scheduled exchange, and explicit fallback re-alignment.
- We separate first-contact opportunity from scheduled energy availability. The first analysis includes role probability and listening duty cycle; the second gives a miss-tail bound under explicit conditional quantile coverage without temporal independence.
- We extend pairwise schedules to single-hop All-to-One collection through versioned discovery, idempotent range assignment, and recovery after clock or link-state loss, and evaluate its slotted-Aloha contention envelope.
- We separate evidence from an STM32WL33 mechanism prototype, independently censored trace-parameterized simulations, and a collision-domain MAC study. Results include seed-block confidence intervals and explicitly bounded claims.

## II. RELATED WORK

### A. Intermittent Computing

Checkpointing systems save volatile state before power loss and restore it after reboot [6], [8], [16]–[18]; task-based runtimes instead divide a program into safely repeatable units [7], [9], [19]. Later work handles peripheral state [20], verification [21], and complete applications [22], [23]. These systems preserve computation on one node. MagPie addresses the complementary problem of acquiring and retaining timing state shared by multiple intermittently powered nodes.

Energy-harvesting wireless sensor networks traditionally use rechargeable batteries or large buffers and schedule sleep over much longer intervals. Duty-cycling [24]–[26] and energy-aware routing [27]–[29] generally assume that a sleeping node retains state and can wake at a predetermined time. MagPie instead targets main-domain brownouts and short active windows.

### B. Wake-Up Radios

Wake-up radios offload idle listening to a receiver drawing microwatts rather than milliwatts [30]–[33]. Greentooth combines WuR and TDMA around a powered base station [34]. CATO is the closest batteryless WuR design: its BEWARE-MAC supports addressed exchange and WEND builds broadcast neighbor discovery [15]. MagPie addresses a different state boundary—retaining relative time when the MCU and main radio repeatedly lose power—and adds energy-gated scheduled collection. CATO is therefore discussed as complementary rather than dismissed or treated as a numerically interchangeable baseline. A direct comparison would require porting both stacks to the same WuR, capacitor, and harvester; the present clock-loss stress test isolates only the persistence mechanism and is not a CATO reimplementation.

### C. Battery-Free Rendezvous and Networking

Classical asynchronous neighbor-discovery protocols illuminate the duty-cycle tradeoff. Disco uses co-prime wake schedules, U-Connect combines symmetric and asymmetric schedules, and Searchlight systematically probes slots [35]–[37]. They provide deterministic or bounded rendezvous under retained clocks and conventional sleep, assumptions that fail when the timing domain itself browns out. MagPie borrows their lesson that schedule density must be priced against discovery delay, but first acquires and then protects a cross-brownout phase. It does not supersede their mobility or heterogeneous-duty-cycle results.

Find uses optimized random delays to discover a peer without infrastructure, and Flync accelerates it when both endpoints observe mains-frequency light flicker [4]. Bonito learns charging-time distributions online and schedules later exchanges from the learned model [5]. These methods avoid a powered coordinator, but their rendezvous quality depends on a charging model or an environmental timing signal.

Coordinator-based systems take the opposite approach. Swift broadcasts synchronization signals from a powered coordinator [12]; FreeBeacon uses coprime cycles to reduce repeated misses after failures [14]; and Pulsar demonstrates scheduled aggregation [13]. Camaroptera and Kingdom likewise use powered infrastructure [38], [39], whereas RICS assumes comparatively stable energy when routing through an intermittent network [40]. Hybrid light/backscatter and other backscatter networks move cost to the physical layer or reader infrastructure [41], [42]; they remain relevant alternatives rather than drop-in protocol baselines. A powered schedule can outperform MagPie, as the Stairs result shows. The distinction is architectural rather than universal performance: MagPie retains a schedule without an always-on control anchor. Its network contribution is versioned single-hop collection, not general topology control or multi-hop routing.

## III. MagPie System Design

We first state the device model and the hardware primitives the design rests on (§III-A) and the multi-threshold energy mechanism governing every operation (§III-B), then detail pairwise communication (§III-C–§III-D) and the networking protocol (§III-E). §III-F quantifies the resulting overhead.

### A. Preliminaries: Device Model and Hardware Primitives

*1) Device and Energy Model:* A battery-free device comprises an energy harvester, a storage capacitor $C$, a microcontroller (MCU), a main radio, a wake-up radio, nonvolatile memory, and a low-power real-time clock. Let $V(t)$ be the capacitor voltage: the main domain is *active* while $V(t)$ exceeds the shutdown threshold $V_{off}$ and *charging* otherwise. Write $X_{i,k}$ for device $i$'s charging time in cycle $k$, from $V_{off}$ to the boot threshold. We make no parametric assumption about its distribution. The analysis in §III-D2 states the weaker conditional coverage property it needs; when that property stops holding, the protocol detects repeated misses and re-estimates its cycle. Devices are stationary, uniquely identified, and within single-hop range.

*2) Primitive 1: Capacitor Voltage Thresholding:* The energy state machine of §III-B needs the MCU to detect when $V(t)$ crosses each of five thresholds. The STM32WL33 provides an analog comparator, a 6-bit DAC, an ADC, and a programmable voltage detector [43]. The proposed power path senses the storage node through a switchable divider, references the comparator from the DAC, and arms only the next expected crossing; an interrupt then selects the following level in Eq. (6). This path is a design specification, not an end-to-end integrated measurement on the present prototype. Divider, comparator, regulator, and reference currents are consequently parameters in the sizing rule below rather than measured MagPie savings.

*3) Primitive 2: Non-Volatile State:* Protocol state that must outlive a main-domain failure comprises the LP-RTC snapshot, peer offsets, synchronization cycle and epoch, aggregation slot, membership version, and neighbor table $N_i$. Timing-critical state fits in the AM1805's 256-byte backup RAM, which shares the separately backed clock domain [44]; stable identifiers and topology versions are mirrored to MCU flash only when they change, not once per charge cycle. A topology update is committed after the state transition and before the $V_{radio\text{-}off}$–$V_{off}$ reserve is consumed. If the RTC backup domain itself fails, its oscillator-failure flag invalidates the backup-RAM record, so stale offsets are never trusted. We do not checkpoint application state, and we do not claim measured NVM-write energy; MagPie composes with existing intermittent-computing runtimes [7], [9], [10].

*4) Primitive 3: The Persistent LP-RTC:* Clock persistence is the load-bearing hardware assumption of MagPie. An oscillator sharing the radio's storage capacitor would stop when that capacitor drains and invalidate every cached offset. We therefore place the LP-RTC in a *separate always-on power domain*: an AM1805 crystal oscillator and counter drawing approximately $I_{rtc}$ = 55 nA in crystal mode, isolated from the main storage node and backed through its dedicated VBAT input [44]. The main domain may brown out while the clock continues, provided the backup reservoir has charge.

**How long the clock survives.** Let $C_{rtc}$ be the reservoir capacitance and $\Delta V_{rtc}$ the voltage headroom between $V_{off}$ and the oscillator's own brown-out level. Ignoring the harvester, the clock domain survives an outage of duration

$$T_{hold} = \frac{C_{rtc} \cdot \Delta V_{rtc}}{I_{rtc} + I_{leak}}, \tag{2}$$

where $I_{leak}$ includes reservoir, isolation, and board leakage. The backup reservoir is therefore sized independently of the main capacitor. Ignoring leakage, 10 mF and 1.5 V of usable headroom yield 75.8 hours (3.2 days) at 55 nA. That value is an ideal upper bound, not a prototype measurement. Leakage of 0.1, 0.5, 1, and 5 $\mu$A reduces the same example to 26.9, 7.5, 4.0, and 0.82 hours, respectively. Temperature, component tolerance, and dielectric leakage can therefore reverse the apparent benefit of a large reservoir. An outage exceeding $T_{hold}$ invalidates cached offsets; MagPie detects that condition (§III-C3) and performs a fresh alignment. Section IV-D4 quantifies only the protocol consequence of loss, not a measured failure distribution.

**How fast the clock drifts.** Even while running, two crystals disagree. If each oscillator has fractional frequency error bounded by $\rho$, their relative offset after an interval $T$ since the last alignment is at most $2\rho T$. A scheduled overlap therefore remains valid only while

$$2\rho T \;\le\; T_{guard} \quad\Longrightarrow\quad \tau_{stale} \;=\; \frac{T_{guard}}{2\rho}, \tag{3}$$

which fixes the staleness threshold at which cached offsets must be discarded. For $\rho$ = 20ppm and $T_{guard}$ = 10ms, Eq. (3)

gives $\tau_{stale} = 250$ s. Here $\rho$ must be the worst-case error over the deployment's temperature range, not a room-temperature typical value. Wake-response jitter and timestamp quantization consume part of $T_{guard}$ at $T = 0$; only the remaining margin covers drift. Increasing the guard extends that interval but raises receiver energy in every exchange; §III-F quantifies the tradeoff.

*5) Sizing the Storage Capacitor:* Each protocol phase $\phi$ starts only when its assigned voltage interval $[V_{l,\phi}, V_{h,\phi}]$ can fund the load, conversion loss, state commit, and quiescent draw:

$$E_{req,\phi} = E_{load,\phi}/\eta_\phi + E_{commit,\phi} + \overline{V}_\phi I_{q,\phi} T_\phi + E_{m,\phi},$$
$$C \geq \max_\phi \frac{2E_{req,\phi}}{V_{h,\phi}^2 - V_{l,\phi}^2}. \quad (4)$$

Here $\eta_\phi$ is conversion efficiency, $I_{q,\phi}$ aggregates static current over duration $T_\phi$, and $\overline{V}_\phi$ is the corresponding average rail voltage. This phase-wise rule is stricter and more useful than assigning an entire handshake to one window, but it becomes predictive only after every term is measured on the target board. The prototype uses 1000 $\mu$F; the simulator's 47 $\mu$F value normalizes Bonito charging times and is not evidence that the prototype can execute its full exchange from that capacitor.

Capacitance is not a free performance improvement. Under an ideal constant charging current, both stored energy and the charge time over a fixed voltage interval scale linearly with $C$; the prototype's 1000 $\mu$F buffer therefore takes 21.3× as long to traverse that interval as 47 $\mu$F. Larger electric-double - layer capacitors also introduce leakage, aging, and cold-start delay. We consequently use equal 47 $\mu$F normalization in every simulated protocol comparison and treat the 1000 $\mu$F testbed only as mechanism validation, not as evidence of an energy or latency advantage over small-buffer baselines.

The WuR listens while the main store is charging. If $I_{harv} > I_{WuR} + I_{leak}$, it reduces net charging current and stretches a charge interval approximately by $(1 - I_{WuR}/(I_{harv} - I_{leak}))^{-1}$. At or below that break-even current, continuous listening would stall charging, so the implementation duty-cycles the WuR. Thus widening $T_{listen}$ is inexpensive only when the local power budget supports it.

Equation (1) also gives an implementable tuning rule. Let $q(L) = \mathbb{E}[\min\{1, (L + w)/X\}]$, estimated from the recent charge history, and let $E_0$ be fixed energy per opportunity. Under the renewal approximation, expected listening-side energy per successful first contact is proportional to

$$J(L, K, \delta) = \frac{E_0 + P_{WuR}\delta L}{p_{role}(K)\,\delta\, q(L)}. \quad (5)$$

A device can search its allowed $(L,\delta)$ values and choose the smallest $J$ subject to a latency constraint; $K = 2$ maximizes pairwise role separation, while dense component discovery instead uses $K \approx m$ for $m$ contenders. This is a first-order controller because the present prototype has no end-to-end current trace from which to calibrate $E_0$.

## B. Multi-Threshold Energy Decision Mechanism

MagPie spans two radio subsystems—the WuR for low-power coordination and the main radio for data exchange—with vastly different energy requirements. To reserve the characterized energy for each committed operation, we partition the capacitor voltage range into five ordered thresholds (Table II):

TABLE II
VOLTAGE THRESHOLDS DEFINING THE ENERGY STATE MACHINE. EACH THRESHOLD GATES A CLASS OF OPERATIONS, ENSURING SUFFICIENT ENERGY BEFORE COMMITMENT; THE SAME FIVE LEVELS ARE DRAWN IN FIGURE 3.

| Threshold | Description |
|---|---|
| $V_{WuR\text{-}2x}$ | Boot threshold; device initiates WuR TX or enters WuR-RX |
| $V_{WuR\text{-}1x}$ | WuR lower bound; proceed to main radio, continue WuR, or recharge |
| $V_{radio\text{-}on}$ | Main radio activation threshold |
| $V_{radio\text{-}off}$ | Radio communication termination threshold; NVM commit point |
| $V_{off}$ | Critical shutdown; device powers off, LP-RTC domain continues |

$$V_{WuR\text{-}2x} > V_{WuR\text{-}1x} > V_{radio\text{-}on} > V_{radio\text{-}off} > V_{off} \quad (6)$$

This provides staged decision-making and an energy reserve: a radio phase starts only after the voltage interval assigned by Eq. (4) is available. The margin absorbs characterized converter and load variation; it is not a guarantee outside the measured operating range.

The energy budget between adjacent thresholds is dimensioned as follows. The region $[V_{WuR\text{-}1x}, V_{WuR\text{-}2x}]$ reserves energy for one WuR operation (either $Radio_{WuR\text{-}tx}$ or $WuR_{rx}$). The region $[V_{radio\text{-}on}, V_{WuR\text{-}1x}]$ reserves a second WuR operation, providing a two-step coordination budget before main-radio engagement. The region $[V_{radio\text{-}off}, V_{radio\text{-}on}]$ sustains one complete main-radio transaction (a TX followed by an RX, or the reverse). Below $V_{radio\text{-}off}$ the device commits state to NVM and sleeps until $V_{off}$.

This creates an asymmetry favoring rapid recovery: a device whose WuR handshake fails to reach a main-radio link falls only to $V_{radio\text{-}on}$ and recovers two regions, whereas a completed exchange falls to $V_{radio\text{-}off}$ and recovers three. Failed coordination is therefore cheaper than successful communication, producing a *listen–charge–listen* pattern that maximizes the fraction of time a device can hear a WuR beacon.

## C. Pairwise Synchronization

We now describe how two battery-free devices discover each other and align their clocks. The protocol proceeds in three stages: role selection, clock alignment via WuR handshake, and persistence of the resulting alignment across power failures.

*1) Role Selection Under Identical Conditions:* A rendezvous requires exactly one transmitter and at least one listener. A naive rule—transmit whenever the LP-RTC counter reaches a multiple of $K$—fails when co-deployed devices advance in lockstep and repeatedly transmit together. MagPie instead derives the decision from identity. At each clock edge, device $i$ with counter $c_i$ transmits a WuR beacon if and only if

$$H(ID_i \,\|\, c_i) \equiv 0 \pmod K, \quad (7)$$

where $H$ is a lightweight hash, and otherwise listens in WuR-RX. Each eligible device therefore transmits on a pseudo-random $1/K$ fraction of edges. Unique identifiers decorrelate roles even with identical energy histories, and re-evaluation prevents one collision from becoming permanent. It does not by itself randomize a periodic wake phase, which is why repeated misses invalidate the alignment premise rather than proving peer absence.

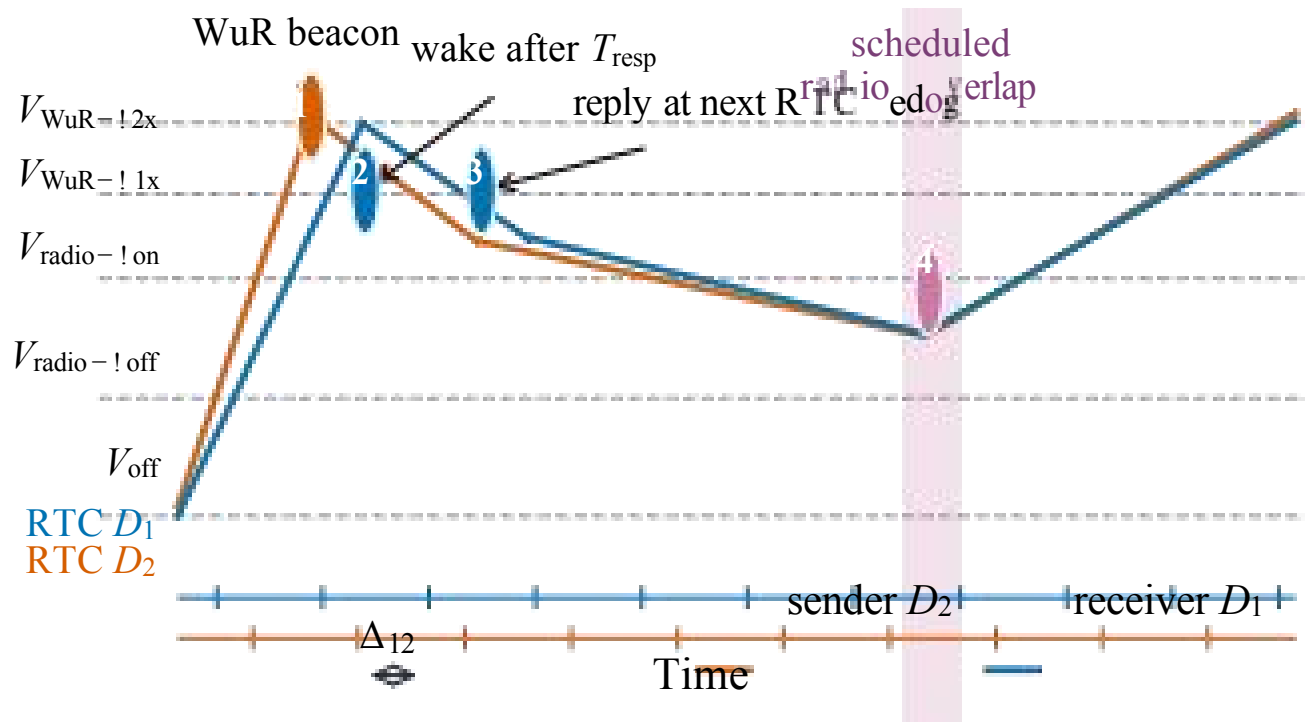

Fig. 3. Clock alignment. (1) $D_2$ sends a WuR beacon; (2) $D_1$ wakes after $T_{resp}$ and measures phase; (3) $D_1$ replies at its next LP-RTC edge; and (4) both devices later open the main radio in the guarded overlap. Dashed lines are the five thresholds in Table II.

**Collision behavior.** With $m$ eligible contenders in one collision domain, the probability of exactly one transmission in an opportunity is

$$P_{one} = \frac{m}{K}\left(1 - \frac{1}{K}\right)^{m-1}, \tag{8}$$

maximized near $K = m$. The choice of $K$ therefore trades idle opportunities against collisions. A failed attempt receives no handshake reply and is retried after a fresh identifier–counter hash; no persistent backoff state is needed. Pairwise simulation uses $K = 2$, and the six-device firmware conservatively uses $K = 32$. For network formation, each component nominates one representative from its current roster and sets $K$ from the advertised component count; a new epoch re-elects representatives. Section IV-D 1 evaluates this slotted-Aloha rule separately from the energy simulator, including fixed $K$, false wakes, and $N = 120$. That model assumes one collision domain and no capture; hidden terminals remain a multi-hop issue.

*2) Clock Alignment via WuR Handshake:* The purpose of alignment is to estimate the relative LP-RTC phase

$$\Delta_{ij} = \phi_i - \phi_j \pmod{T_{clk}} \tag{9}$$

between two devices, so that each can compute the other's clock edges thereafter.

Figure 3 illustrates the procedure. Suppose Eq. (7) selects $D_2$ as sender while $D_1$ listens in WuR-RX:

*a)* $D_2$ transmits a WuR beacon ($Radio_{WuR\text{-}tx}$) carrying its current LP-RTC counter value $c_2$, and records its own local transmit instant $t_2^{tx}$.

*b)* $D_1$ detects the beacon and its MCU becomes ready after the nominal wake-up response time $T_{resp}$, dominated by oscillator stabilization. Let $t_1^{rx}$ be $D_1$'s local time at that moment and let $T_{sleep}$ be the residual interval until $D_1$'s next clock edge. Using the calibrated mean response, $D_1$ recovers the phase difference as

$$\Delta_{12} = \left(t_1^{rx} - T_{resp} - c_2 T_{clk}\right) \bmod T_{clk}, \tag{10}$$

and equivalently obtains $T_{sleep} = T_{clk} - \Delta_{12}$ as the wait to its own next edge. The arithmetic is modulo $T_{clk}$, so $T_{resp}$ may span more than one tick; its measured jitter, rather than its mean, contributes to $T_{guard}$.

*c)* $D_1$ replies with a WuR beacon at its next clock edge, carrying $c_1$ and the measured $\Delta_{12}$. This lets $D_2$ compute the reverse offset $\Delta_{21} = -\Delta_{12} \bmod T_{clk}$ and confirms to $D_1$ that the link is bidirectional.

After this exchange both devices hold each other's phase and can compute the other's clock edges without further measurement. The residual error in Eq. (10) is bounded by the LP-RTC resolution plus $T_{resp}$ jitter—the quantity $T_{guard}$ must cover at $T = 0$, growing thereafter at $2\rho$ per unit time.

*3) Shutdown, Recovery, and Staleness Detection:* When $V(t)$ falls below $V_{radio\text{-}off}$, the device commits the state enumerated in §III-A to NVM and sleeps; below $V_{off}$ it powers down entirely, leaving only the LP-RTC domain alive. On recharging to $V_{WuR\text{-}2x}$ it reloads that state and resumes from the last committed point. Because the LP-RTC has continued counting, the stored offsets are still meaningful: the device reads the current counter and computes the next scheduled rendezvous directly from $\Delta_{ij}$. A single successful alignment therefore persists across an arbitrary number of power failures, provided two conditions hold.

The first condition is that the clock domain survived, *i.e.* the outage was shorter than $T_{hold}$ of Eq. (2). At boot, firmware checks the AM1805 oscillator-failure flag, which is asserted after initial power-up or oscillator failure [44]. A set flag invalidates all cached offsets; firmware clears it only after a successful alignment.

The second condition is that accumulated drift has not exceeded the guard interval. The device compares the current counter against the stored snapshot; if the elapsed time exceeds $\tau_{stale}$ from Eq. (3), it invalidates its cached offsets and re-enters alignment. Both conditions are checked at every boot, cost a handful of instructions, and fail safe: a device that wrongly believes its offsets are valid loses at most $R_{\max}$ cycles before the fallback of § III-D4 triggers a full re-alignment.

## D. Sustained Connection

Once aligned, a pair exchanges data through scheduled radio overlap rather than repeated handshakes.

*1) Sizing the Synchronization Cycle:* Both devices must agree on how long to wait between communication opportunities. Let $X_1, X_2$ denote their charging times, and let $\hat{q}_i(\alpha)$ be device $i$'s empirical $\alpha$-quantile over its last $W$ cycles ($W = 32$). During alignment each device transmits $\hat{q}_i(\alpha)$, and both adopt

$$T_{cyc} = \left\lceil \max\{\hat{q}_1(\alpha), \hat{q}_2(\alpha)\}/T_{clk} \right\rceil \cdot T_{clk}. \tag{11}$$

The estimate is a quantile, not a maximum: it deliberately accepts nominal miss probability $1 - \alpha$ instead of pretending that the next sample is bounded by the past. We use $\alpha = 0.9$. More generally, the distribution-free attempt bound below suggests choosing

$$\alpha^* \in \arg\min_{\alpha > 1/2} \frac{\max_i \hat{q}_i(\alpha)}{2\alpha - 1}, \tag{12}$$

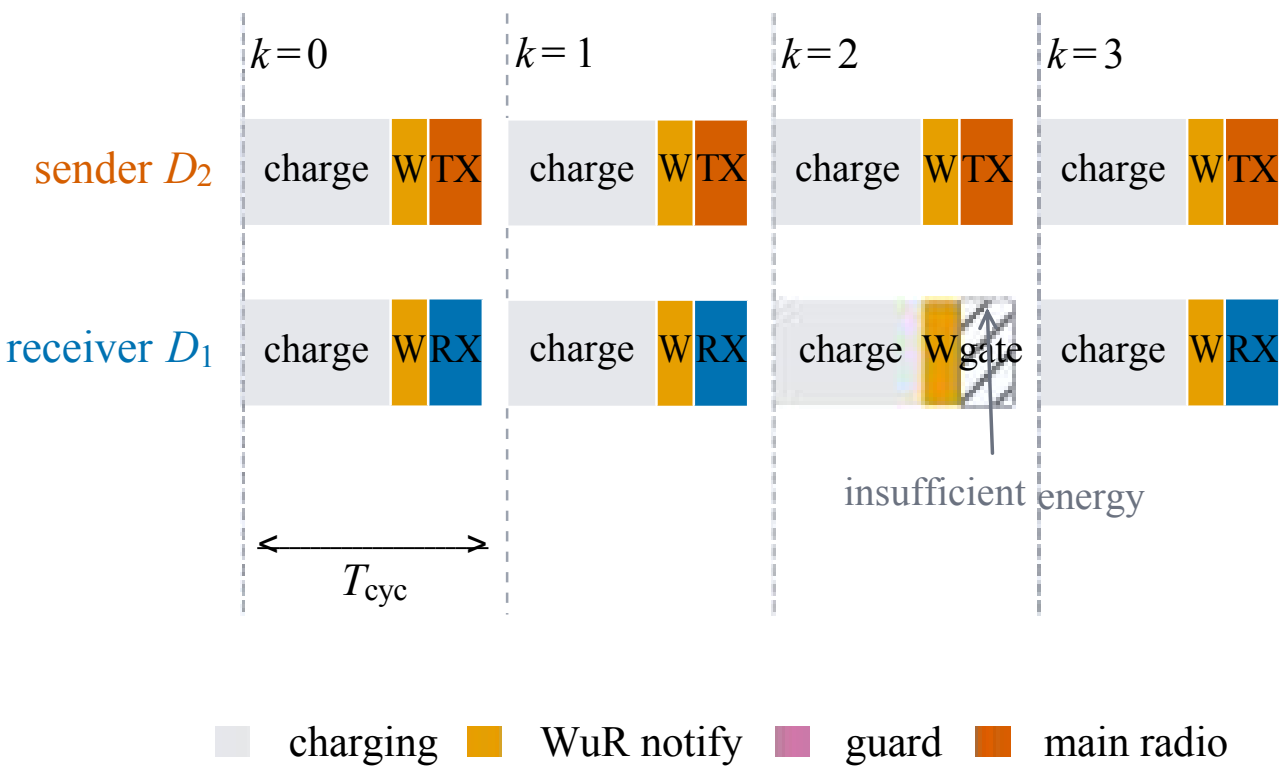

Fig. 4. Sustained pairwise connection. A sender notification (W) precedes each guarded main-radio overlap. At $k = 2$ the receiver's energy gate blocks its radio; the persistent schedule is retained and the pair retries at $k = 3$.

subject to the application's required miss tail. This prices a longer cycle against fewer expected attempts. The current firmware fixes $\alpha$ rather than running this controller. Its rolling empirical quantile also does not guarantee conditional coverage under arbitrary drift; an online calibration test must treat a sustained excess miss rate as a regime change and advance the schedule epoch.

*2) Rendezvous Analysis:* We can now bound what MagPie guarantees. Equation (1) concerns whether a beacon can be heard; the remaining uncertainty is whether both endpoints hold enough energy at the scheduled boundary. Fix $T_{cyc}$ and let $\mathcal{F}_{k-1}$ contain the charging history before cycle $k$. The estimator is *conditionally calibrated* at level $\alpha$ when, for each endpoint $i$ and cycle $k$,

$$\Pr[X_{i,k} \le T_{cyc} \mid \mathcal{F}_{k-1}] \ge \alpha. \tag{13}$$

This explicitly rules out an unobserved regime shift that makes the selected cycle systematically too short. Under Eq. (13), the conditional probability that both endpoints are ready is bounded by the Fréchet–Hoeffding inequality:

$$p_k \ge p_0 \triangleq \max(0, 2\alpha - 1). \tag{14}$$

If the two charging processes are conditionally independent, the stronger bound $p_k \ge \alpha^2$ applies.

**Proposition 1** (Rendezvous tail bound). *Under Eqs. (13)–(14), the number of cycles $M$ until the first successful exchange satisfies*

$$\Pr[M > m] \le (1 - p_0)^m, \qquad \mathbb{E}[M] \le \frac{1}{p_0}. \tag{15}$$

*For independent, identically distributed cycles with constant joint success probability $p$, these inequalities reduce to the familiar geometric equalities. At $\alpha = 0.9$, $p_0 = 0.8$, so the expected number of attempts is at most 1.25 and five consecutive misses have probability at most* $3.2 \times 10^{-4}$.

The result is a bound, not determinism. When the conditional coverage premise fails, no tail guarantee follows. Operationally, however, a miss does not erase phase: the next opportunity remains computable from the surviving clock, whereas a prediction-based discovery may have to search for the peer again.

**Cycle updates.** Equation (13) can fail after an environmental change. The sender therefore owns a monotonically increasing schedule epoch and carries $(epoch, T_{cyc}, t_{next})$ in each WuR notification. It recomputes Eq. (11) from its rolling window and the peer's last advertised estimate; the receiver adopts only newer epochs. After $R_{\max}$ receiver-observed misses, that receiver initiates the full-handshake fallback of §III-D4, during which the sender transfers a fresh estimate and schedule. This explicit versioning avoids assuming that an unpowered receiver and a one-way sender observe the same miss sequence.

*3) Acknowledgment-Free Scheduled Activation:* A full alignment uses a return WuR beacon to measure both directions. Once phase is known, scheduled activation omits that *WuR* acknowledgment (Figure 4). The sender emits a notification at the cycle boundary and schedules its main-radio packet to coincide with the receiver's wake-up, computed from $\Delta_{ij}$ and $T_{resp}$. The receiver opens its radio $T_{guard}$ early to absorb residual error. Both endpoints verify $V(t) > V_{radio\text{-}on}$ before main-radio activation; an energy-starved endpoint simply remains absent.

This optimization must not be confused with end-to-end reliable delivery. In a one-way collection slot, the sender receives no data acknowledgment and therefore cannot infer that the sink accepted its packet. The simulator reports receiver-side receptions, and the testbed reports packets logged at the sink. An application requiring confirmation must add a later cumulative bitmap or a short main-radio ACK and include its energy in Eq. (4); neither mechanism is implemented or evaluated here. "Acknowledgment-free" in this paper consequently describes scheduled activation, not a reliability guarantee.

*4) Failure Recovery:* A receiver that was energy-qualified and heard a notification but observed no scheduled payload can count a miss; possible causes include sender energy, payload loss, drift, or a stale cycle. A one-way sender cannot make the symmetric observation. An isolated receiver-side miss leaves the schedule intact. Equation (3) prevents known drift from exceeding $T_{guard}$, while the epoch mechanism communicates cycle changes whenever the receiver is reachable.

After $R_{\max}$ observable consecutive misses ($R_{\max} = 3$), the observing endpoint initiates a full bidirectional WuR handshake and remeasures phase and cycle. In All-to-One collection this endpoint is normally the sink; if it is itself unpowered, recovery cannot start until it returns. A smaller threshold performs more needless alignments; a larger one can leave the pair disconnected for $R_{\max}T_{cyc}$. The evaluation fixes this parameter and models observable misses; it does not claim end-to-end acknowledgment semantics or an experimentally optimal value.

## E. Networking Protocol

We now extend pairwise communication to $N$ reporting battery-free devices and one logically distinguished sink in a static, single-hop radio domain. Each reporter transmits in a distinct slot of a shared *distribution cycle* $C_{dis}$. The sink does not allocate indices and need not be externally powered; however, a collection round can begin only when it has enough energy to receive that round. The simulator idealizes this gate as always true, an explicit limitation in §IV-A. Construction has two phases: *component discovery*, which creates a versioned timing tree, and *slot distribution*, which allocates a unique index to every reporter.

TABLE III
LOGICAL MESSAGES AND TRANSPORT. "PERSISTENT EFFECT" IS COMMITTED BEFORE THE RESERVED SHUTDOWN MARGIN IS CONSUMED.

| Object | Transport | Core fields | Persistent effect |
|---|---|---|---|
| Align | WuR | ID, counter | peer phase |
| Notify | WuR | epoch, next tick | cycle update |
| Roster | main radio | comp. ID, epoch, bitmap | membership |
| Range | main radio | epoch, interval, slot | slot assignment |
| Data | main radio | epoch, source, payload | sink reception |

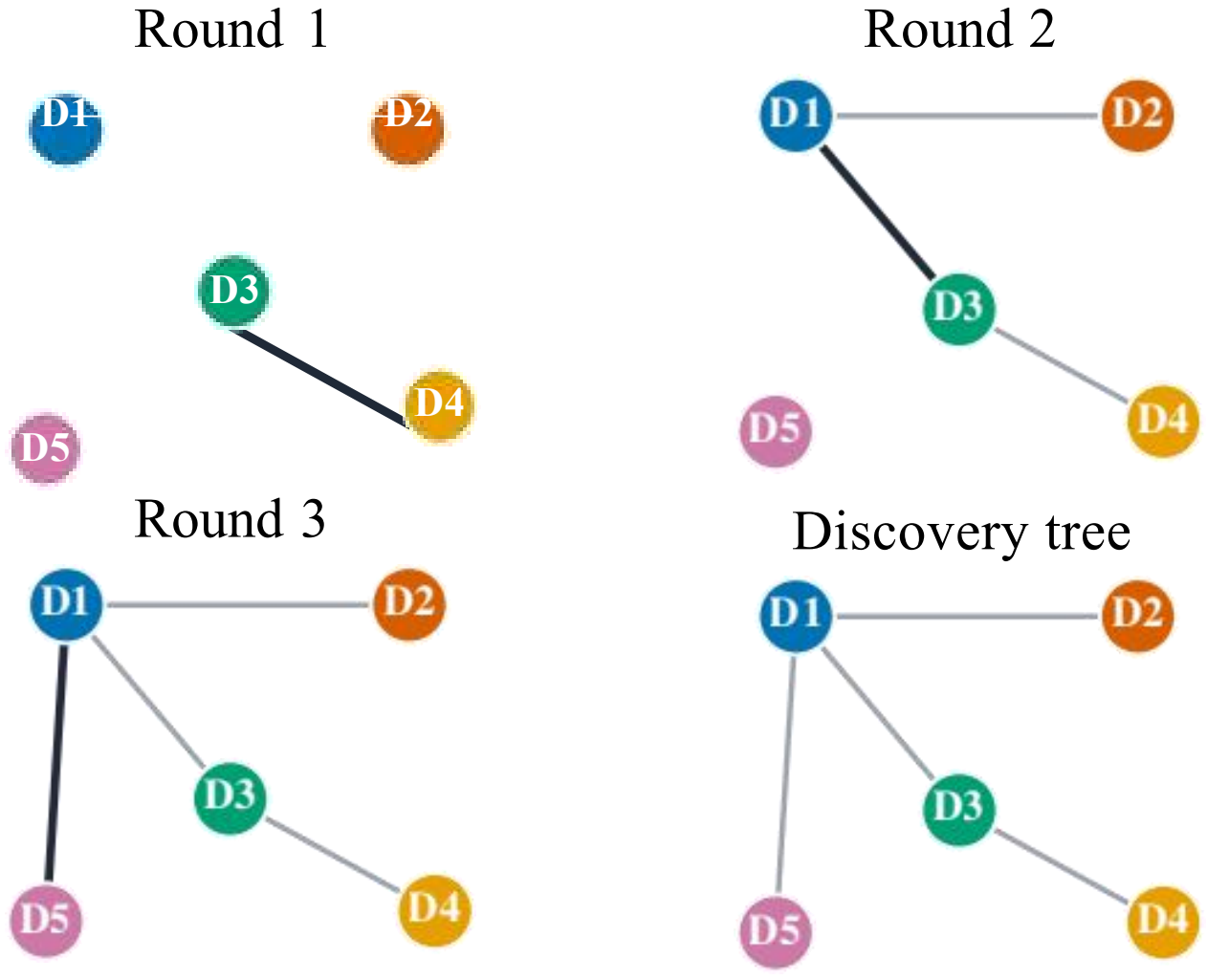

Fig. 5. Logical component merges for five reporters. Only inter-component contacts are retained, producing a tree with $N-1$ timing edges. In one collision domain, at most one solo WuR transmission succeeds per contention opportunity.

Table III separates the logical control objects and their placement. The 15-byte bitmap at $N = 120$ is a main-radio roster payload after WuR alignment, not a WuR wake address. The evaluation fixes a 32-byte main-radio packet, so that bitmap leaves 17 bytes for the component identifier, epoch, and link-layer header in the modeled frame. Exact WuR serialization and link-layer overhead are platform specific and are not included in the simulator's primitive-energy table.

*1) Phase 1: Component Discovery:* Figure 5 illustrates one five-reporter execution. Each component carries an exact membership bitmap and epoch, and nominates one identifier-hashed representative per contention epoch. Representatives apply Eq. (7) with $K$ matched to the estimated number of components. A solo beacon can align with a listening representative from another component; two or more concurrent beacons collide in the single-domain model. A successful inter-component contact unions the rosters, advances the epoch, and becomes a merge-tree edge stored at both endpoints. Intra-component contacts are ignored. Discovery completes when the merged bitmap contains the *provisioned* identifier universe of size $N$. This rule yields exactly $N - 1$ accepted edges, but it does not discover an unknown population; dynamic join and leave require a separate membership service.

**Discovery is amortized.** Because $N_i$, phase offsets, membership, and slot assignment survive main-domain brownouts, discovery is not repeated per packet. A device re-enters it only when an offset becomes stale, the RTC failure flag is set, or a membership epoch changes.

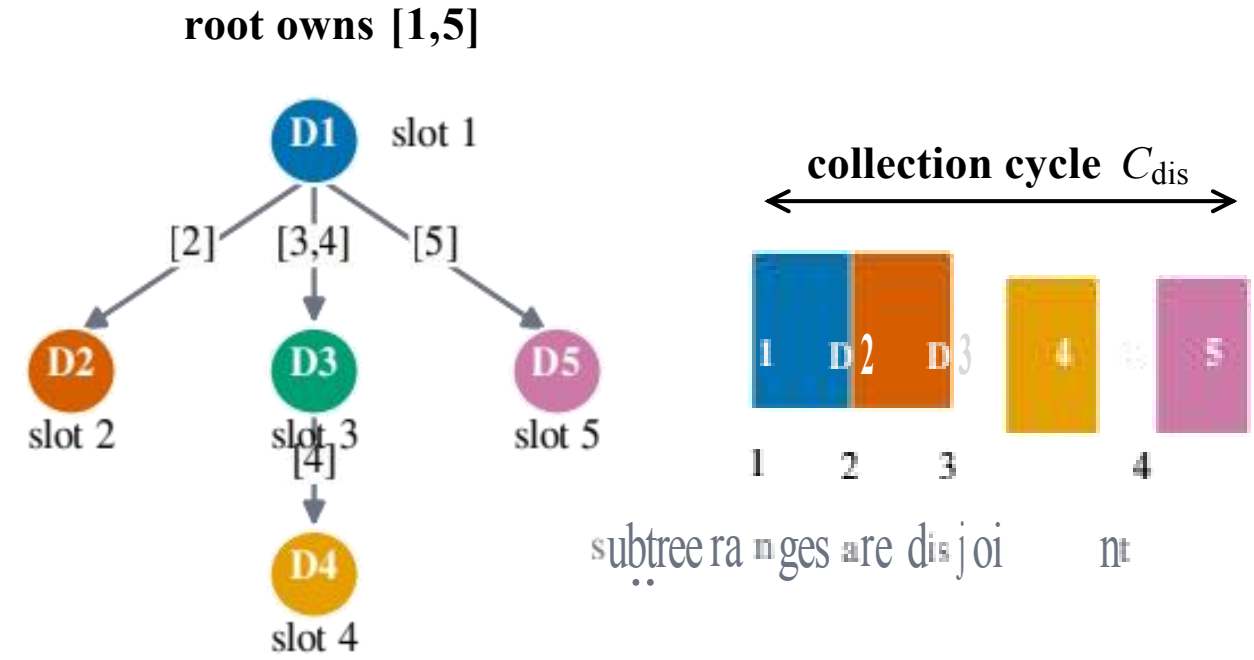

Fig 6 Slot distribution for five reporters Root $D_1$ owns [1, 5], retains slot 1, and partitions the remaining interval into disjoint subtree ranges A reporter stores (*epoch*, *range*, *slot*) before acknowledging; the final cycle is therefore unique despite message replay.

*2) Phase 2: Slot Distribution:* We set $C_{dis} = NT_{clk}$. The smallest identifier roots the merge tree. An upward pass first reports each subtree's size; a downward pass gives the root $[1, N]$ and recursively partitions it into contiguous child ranges in identifier order. Each node takes the first index of its range and divides the remainder according to the reported subtree sizes. Figure 6 shows the five-reporter case.

**Why replay cannot create duplicate slots.** Every assignment carries the component epoch and a closed integer range. A child writes (*epoch*, *range*, *slot*) to persistent state before acknowledging. Repeating a message after either endpoint browns out is therefore idempotent: an equal epoch and range reproduce the same decision, while an older epoch is discarded. Because sibling ranges are disjoint by construction, concurrent branches cannot issue the same index. A missing acknowledgement leaves a gap until the identical assignment is replayed; it does not release the range for reassignment within that epoch. After all confirmations, the root propagates (*epoch*, $C_{dis}$, $t_0$) and the sink adopts the same epoch and origin through its pairwise schedule with the root.

*3) Maintenance Under Clock or Link-State Loss:* Each tree edge is `valid`, `suspect`, or `expired`. $R_{\max}$ observable misses make it suspect and trigger the full handshake of §III-D4; a successful handshake refreshes its offset and returns it to valid. If only the clock domain is lost while the topology epoch survives in flash, the node invalidates all adjacent offsets and re-aligns those edges before using its slot; it does not silently reuse their timestamps. After $L_{\max} = 3$ complete handshake timeouts, the endpoints expire the edge, increment the component epoch, and the detached subtree re-enters discovery. A parent outage can therefore pause its subtree but does not change siblings' epochs until the edge expires. The static, error-free evaluation does not validate these timers under mobility or fading.

**Control and storage complexity.** A completed merge tree has $N - 1$ accepted alignments. The two slot passes use at most $2(N-1)$ successful parent–child transfers, and each collection round contains $N$ reporter transmissions. Per-node persistent storage is $O(d_i)$ for $d_i$ tree neighbors and total topology storage is $O(N)$. Candidate selection can inspect $O(N^2)$ pairs in the worst case. Slotted contention adds a random number of idle and collided opportunities quantified in §IV-D1; the separate energy simulator's matching must not be used to infer that cost.

TABLE IV
MODEL-BASED PRIMITIVE COSTS USED BY THE PROTOCOL.

| Phase | Cost | Latency | Frequency |
|---|---|---|---|
| WuR beacon TX | 38.4 $\mu$J | 1.12 ms | per notification |
| WuR listening | 12 $\mu$W | $T_{listen}$ | while enabled |
| Data packet TX | 40.8 $\mu$J | 1.02 ms | per exchange |
| Main-radio RX | 165 $\mu$J | 11.0 ms | payload + guard |

*4) Cycle Reconciliation:* A device must honor both its pairwise cycle $C_{sync}$ and the distribution cycle $C_{dis}$. Rather than take an arbitrarily large least common multiple, it rounds $C_{sync}$ up to the next multiple of $C_{dis}$:

$$C_{op} = \left\lceil \frac{C_{sync}}{C_{dis}} \right\rceil C_{dis}, \tag{16}$$

expressed in $T_{clk}$ ticks and committed with the current epoch. This gives $C_{sync} \leq C_{op} < C_{sync} + C_{dis} \leq 2\max(C_{sync}, C_{dis})$. The added wait is less than one distribution cycle, not one clock tick.

### F. Overhead Analysis

A coordination scheme that consumes the budget it protects is of no use. Table IV therefore exposes the primitive costs used by the energy gate and simulator. Values are computed as measured duration times configured active power; they are not end-to-end shunt measurements and exclude regulator loss and NVM commits.

Excluding WuR listening, conversion loss, and state commits, the modeled active cost of one acknowledgment-free scheduled data transfer is 79.2 $\mu$J at the sender (one notification plus data TX) and 165 $\mu$J at the receiver. Initial alignment adds two WuR beacon transmissions, 76.8 $\mu$J across the pair. Listening itself is not free: 12 $\mu$W for the maximum 30 s window is 360 $\mu$J drawn from the harvester and appears as a longer charge interval through the relation in §III-A5. A false wake additionally invokes platform-dependent MCU start-up and is not represented by these four primitives.

The receiver is the expensive scheduled role: at the configured 15 mW, a 10 ms guard costs 150 $\mu$J before payload reception. Equation (3) is therefore a genuine tradeoff. A wider guard extends offset lifetime but is paid every cycle; periodic re-alignment is preferable once its amortized cost is lower. Because the omitted terms can be comparable to the WuR load, the table supports a first-order design study only. End-to-end shunt measurements of regulator input energy, threshold circuitry, false-wake handling, NVM commits, and backup-domain leakage are required before Eq. (4) can size production hardware or support an energy-efficiency claim.

## IV. Evaluation

We answer six questions: (1) How quickly does MagPie establish a pairwise schedule under right censoring? (2) What is the sustained exchange interval after alignment? (3) How does single-domain WuR contention scale? (4) How do collection and clock loss scale with $N$? (5) How sensitive are the results to cycle error, drift, and retry policy? (6) Which mechanisms execute on intermittently powered hardware?

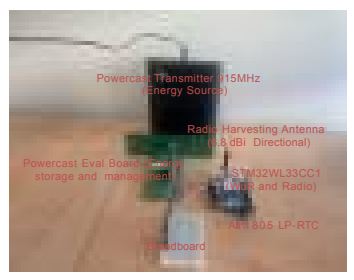
(a) Prototype

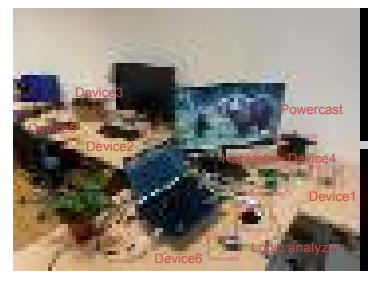
(b) Six devices

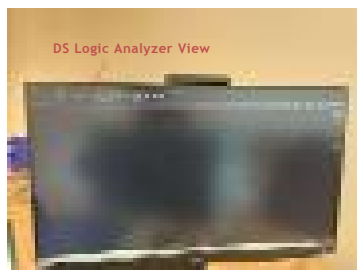

(c) Logic analyzer

Fig. 7. Experimental testbed: (a) prototype, (b) six-device All-to-One deployment, and (c) logic-analyzer setup. Device 6 is the sink.

TABLE V
PROTOTYPE AND SIMULATOR PARAMETERS.

| Parameter | Prototype | Simulator |
|---|---|---|
| WuR listening current | 4 $\mu$A | 12 $\mu$W at 3 V |
| WuR response $T_{resp}$ | 133 ms | 113 ms |
| LP-RTC tick $T_{clk}$ | 100 ms | 30 ms |
| Listening window $T_{listen}$ | up to 30 s | 30 s |
| Role modulus $K$ | 32 | 2 (pairwise); $K \approx m$ (MAC study) |
| Main storage | 1000 $\mu$F | 47 $\mu$F normalization |
| Cycle quantile $\alpha$ | 0.9 | 0.9 |
| Retry threshold $R_{max}$ | 3 | 3 |
| Post-event state | — | 0.9/0.8/0.3 |
| Main-radio TX/RX | 40/15 mW | 40/15 mW |
| Data packet | 32 B | 32 B |

### A. Experimental Setup

*1) Prototype:* The prototype uses the STM32WL33CC1, which integrates a Cortex-M0+ MCU, sub-GHz transceiver, and low-power wake-up receiver [43]. The WuR draws approximately 4 $\mu$A while listening. An AM1805 in the isolated domain of §III-A4 provides the persistent 32.768 kHz time base [44]. A Powercast TX91502/P2110 link supplies controlled 915 MHz energy to a 1000 $\mu$F main buffer. Harvesting and communication are time-multiplexed because they share a band. Here "infrastructure-free" refers to control and scheduling, not to this laboratory energy source; the prototype is not an ambient-harvesting deployment.

Table V separates prototype settings from simulator settings. The pairwise simulator uses a finer tick and $K = 2$; the six-device firmware uses $K = 32$. The separate contention study adapts $K$ to component density. Results from these artifacts are therefore not numerically interchangeable.

*2) Simulation Framework:* The Python discrete-event simulator represents charging intervals, five energy regions, WuR discovery, scheduled exchange, component construction, and slotted collection. It uses five harvesting settings published with Bonito [11]: Office and Stairs (indoor and outdoor light), Washer (machine vibration), Jogging (human motion), and Cars (vehicle vibration). Checked-in experiments use seeded *synthetic traces parameterized from those settings*, not direct HDF5 replay. Exponential, normal, or two-component Gaussian mixtures reproduce scenario-level mean, variance, drift, and device variation. Charging intervals are scaled from the Bonito artifact's 17 $\mu$F reference to 47 $\mu$F by 47/17. The normalized storage state is 1 at WuR-ready; a receive timeout, unanswered WuR transmission, and completed main-radio operation leave

TABLE VI
EVIDENCE MAP. EACH ARTIFACT SUPPORTS ONLY THE CLAIM IN ITS ROW.

| Artifact | Supports | Does not support |
|---|---|---|
| 2-node logic traces | alignment execution and rate | ambient energy or packet delivery |
| 6-device run | formation/slot execution | a delivery ratio or large scale |
| Pairwise simulator | censored relative latency | hardware energy |
| Network energy model | slot/energy bookkeeping | RF contention or sink outage |
| Contention model | single-domain collisions | capture, hidden terminals, energy |

0.9, 0.8, and 0.3. These are disclosed model parameters, not measured voltages.

Pairwise synchronization uses 100 independent first-rendezvous trials per setting, arranged as ten trials in each of ten seed blocks. Each trial is administratively censored at $\tau = 5000$ s. MagPie starts from the modeled post-exchange storage state 0.3, avoiding the zero-charge first event created by an initially full capacitor. We report completion count and restricted mean synchronization time $\mathrm{RMST}(\tau) = \mathsf{E}[\min(T,\tau)]$; 95% intervals are Student-$t$ intervals over the ten seed-block RMSTs. Completed-event CDFs are normalized by all 100 trials, so a curve ending below one exposes censoring.

Baselines are Find and, only in Office, Flync [4]; the other sources provide no mains-frequency optical reference. An optimistic *powered-coordinator reference* captures the scheduled-cycle mechanism shared by FreeBeacon and Pulsar [13], [14]. Its cycle is selected per scenario from 17 candidates between 2 and 5003 ms using the lowest RMST on that scenario. This same-data oracle favors powered infrastructure and is not a feature-complete reimplementation. It also lacks a WuR and therefore is not a causal coordinator ablation: the full $\{$WuR/no WuR$\}\times\{$backed clock/no clock$\}\times$ $\{$coordinator/no coordinator$\}$ hardware matrix remains unmeasured. Sustained connection is compared with a seeded Bonito implementation using five warm-up samples and $\hat{\lambda} = 1/\bar{X}$ [5].

The network energy model runs ten seeds and 50 collection rounds per point. It assumes an error-free single-hop channel and an always-ready sink, and abstracts discovery as a random nonoverlapping matching. Its formation time is therefore discarded from the scalability claim; only energy gating, assigned-slot execution, and recovery bookkeeping are reported. A second, MAC-only experiment replaces that matching with one collision domain and slotted Aloha. Each component representative transmits with probability $1/K$; exactly one transmitter plus a listener merges two components, whereas multiple transmitters collide without capture. Five hundred independent trials report mean, P95, and confidence intervals through $N = 120$. A false wake makes a representative unavailable for one opportunity. This model still excludes hidden terminals, fading, and energy readiness.

### B. Pairwise Synchronization

Figure 8 and Table VII no longer discard timed-out trials. MagPie completes 100/100 in every scenario; Find completes all trials in Office, Stairs, and Washer, 93 in Jogging, and 36 in Cars. Its RMST is 82.7–755.8 $\times$ that of MagPie across the five parameterized scenarios. Office Flync completes all trials with a 38.2 $\times$ RMST ratio. These ratios describe the disclosed simulator, not a universal hardware speedup.

The powered reference exposes the architectural tradeoff. It is 1.3 $\times$ faster than MagPie in stable Stairs; MagPie has lower RMST in the other four settings. Because the reference is tuned on each scenario and changes both the coordinator and WuR factors, the result is a system comparison rather than an attribution of gain to one component.

### C. Sustained Pairwise Connection

After discovery, both protocols complete 100 exchanges in the first four scenarios, while Bonito does not discover a peer in Cars before the horizon. MagPie reduces the mean sustained interval by 1.8–3.6$\times$. The gap is smaller than in first contact because both designs benefit once a shared interval exists. MagPie's distinction is that the interval is retained by the backed clock and a miss triggers a bounded retry rather than immediate rediscovery.

### D. Network Evaluation

*1) Collision-Domain Formation:* Figure 10(a) replaces the nonoverlapping-matching assumption for the MAC question. At $N = 120$, adaptive $K = m$ needs a mean 318.0 opportunities (95% CI 316.1–320.0) and P95 359.0 (bootstrap CI 353.1–362.1). If every 30 ms tick were energy eligible, those values would be 9.54 and 10.77 s; because energy readiness is omitted, they are lower bounds on wall-clock formation. Fixed $K = 32$ raises P95 to 725 opportunities. Panel (b) confirms Eq. (8): $K/m = 1$ is best among the tested ratios, whereas $K/m = 0.5$ and 2 increase mean opportunities from 318.8 to 441.3 and 385.0. In panel (c), making 40% of representatives unavailable per opportunity raises P95 from 355 to 412. These results quantify collision and false-wake sensitivity only; capture, hidden terminals, and spurious-wake energy require measurements and a topology-aware PHY model.

*2) Fixed Size:* Table IX reports $N = 20$ reporters over ten seeds and 50 rounds. With $T_{clk} = 30$ ms, $C_{dis} = 600$ ms. Stable Stairs and Office reach 99.7–99.8% receiver-side delivery; Washer reaches 89.0%. Jogging and Cars fall to approximately 70% because reporters often lack radio-grade energy at their assigned instant. The error-free channel makes these losses energy-related by construction.

*3) Scalability:* Figure 11 scales assigned-slot execution from 6 to 120 reporters. Collection time grows with $C_{dis} = NT_{clk}$: at $N = 120$ it is 3.57 s in Office, Stairs, and Washer, 4.32 s in Jogging, and 11.46 s in Cars. Mean receiver-side delivery is 99.8%, 99.7%, 88.3%, 70.4%, and 70.2%, respectively. The nearly flat delivery trend shows only that this energy model does not exhaust its assigned slots as $N$ increases. It says nothing about collisions (Figure 10), interference, or an intermittent sink.

*4) LP-RTC Failure Stress Test:* We next force the backed clock to lose state independently with probability $p$ per cycle. This is a protocol stress test, not a measured AM1805 distribution. At $p = 0$, delivery is 99.8% with no re-alignments. At $p = 0.2$, delivery is 80.6%, 79.2%, 80.2%, and 79.5% for $N = 5, 10, 20, 40$; mean re-alignments are 52.5, 111.8, 215.4, and 448.9. The approximately linear recovery work explains why Eq. (2) should be sized conservatively. IID failures do not model a common temperature excursion or shared reservoir fault, which could invalidate many edges together.

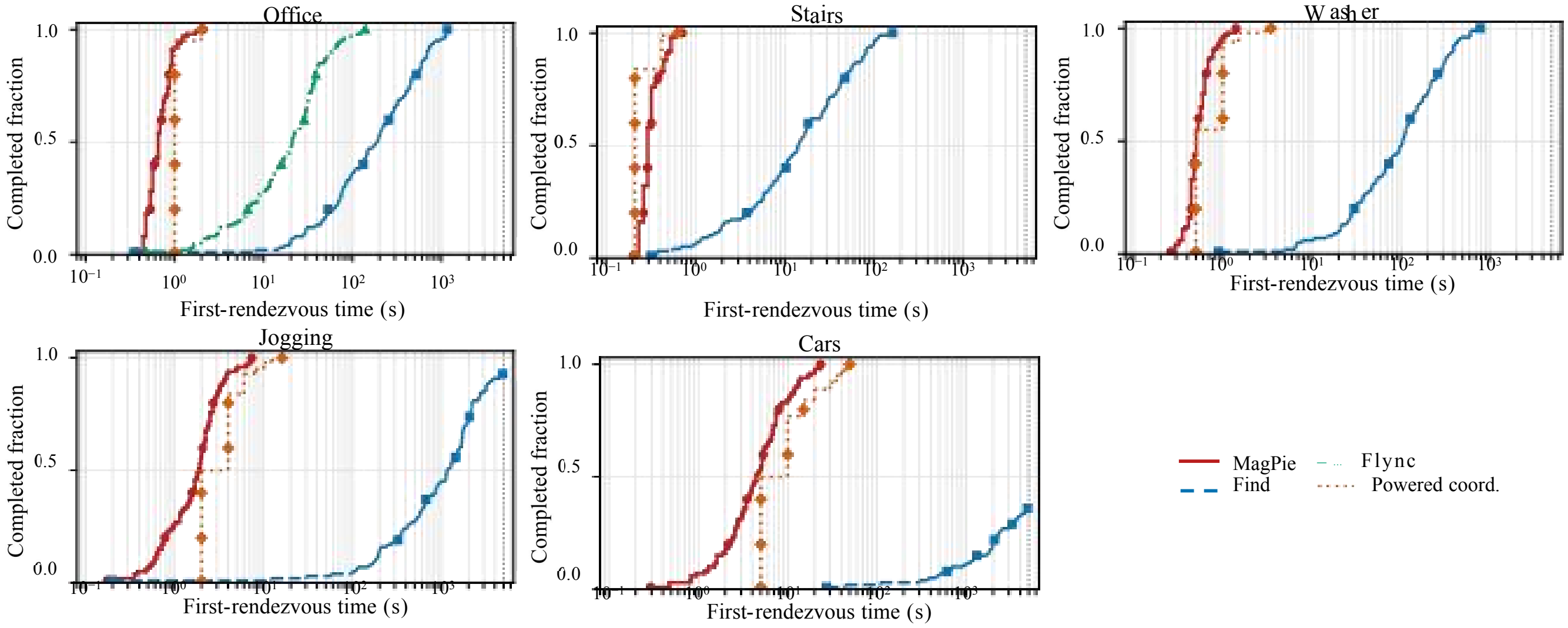

Fig. 8. Empirical completion curves for independent first-rendezvous trials (logarithmic time). Each jump is divided by all 100 trials, so an endpoint below one shows administrative censoring at 5000 s. Flync is physically available only in Office.

TABLE VII
FIRST-RENDEZVOUS RMST IN SECONDS TO $\tau$ = 5000 S. BRACKETS ARE 95% INTERVALS OVER TEN SEED-BLOCK MEANS; THE SECOND LINE IS COMPLETIONS/100. "N/A" DENOTES AN UNAVAILABLE ENVIRONMENTAL TIMING SIGNAL.

| | Office | Stairs | Washer | Jogging | Cars |
|---|---|---|---|---|---|
| MagPie | 0.71 [0.65, 0.78]<br>100/100 | 0.33 [0.31, 0.34]<br>100/100 | 0.56 [0.52, 0.60]<br>100/100 | 2.04 [1.73, 2.35]<br>100/100 | 5.86 [4.78, 6.94]<br>100/100 |
| Find | 288.57 [215.80, 361.34]<br>100/100 | 27.30 [18.62, 35.99]<br>100/100 | 156.02 [130.98, 181.07]<br>100/100 | 1541.78 [1180.50, 1903.06]<br>93/100 | 3892.44 [3513.67, 4271.21]<br>36/100 |
| Flync | 27.13 [21.63, 32.64]<br>100/100 | n/a | n/a | n/a | n/a |
| Powered coord. | 1.05 [1.01, 1.08]<br>100/100 | 0.25 [0.24, 0.25]<br>100/100 | 0.79 [0.68, 0.89]<br>100/100 | 3.65 [3.15, 4.14]<br>100/100 | 11.41 [8.60, 14.21]<br>100/100 |

TABLE VIII
SUSTAINED CONNECTION AFTER DISCOVERY. EACH SUCCESSFUL SETTING CONTAINS 100 EXCHANGES; INTERVAL IS THE MEAN BETWEEN EXCHANGES.

| Scenario | MagPie (ms) | Bonito (ms) | Ratio |
|---|---|---|---|
| Office | 951 | 3,258 | 3.43 × |
| Stairs | 215 | 780 | 3.64 × |
| Washer | 652 | 2,174 | 3.34 × |
| Jogging | 3,052 | 5,484 | 1.80 × |
| Cars | 9,479 | — | — |

TABLE IX
NETWORK ENERGY-MODEL RESULTS FOR $N$ = 20 (10 SEEDS, 50 ROUNDS). BRACKETS ARE 95% SEED-LEVEL CONFIDENCE INTERVALS.

| Scenario | Mean collection (s) | Delivery |
|---|---|---|
| Stairs | 0.57 [0.57, 0.57] | 99.74 [99.55, 99.93]% |
| Office | 0.61 [0.60, 0.63] | 99.81 [99.71, 99.91]% |
| Washer | 0.58 [0.57, 0.58] | 88.99 [88.26, 89.72]% |
| Jogging | 2.21 [2.02, 2.40] | 69.63 [69.10, 70.16]% |
| Cars | 6.90 [6.30, 7.50] | 69.97 [69.40, 70.54]% |

## E. Sensitivity and Assumption Checks

**Cycle estimation.** For each seeded scenario we compute Eq. (11) from the previous $W$ = 32 samples and test the next pair of charging times. With no artificial error, observed joint readiness is 84.7–94.1%, above the distribution-free 80% lower bound for a conditionally calibrated $\alpha$ = 0.9. A 25% underestimate reduces readiness to 25.4–74.5%; a 25% overestimate raises it to 91.4–99.9% but lengthens every scheduled cycle by approximately 25%. Thus overestimation spends latency, whereas systematic underestimation can invalidate the operating point and trigger epoch updates. This sweep uses the same seeded synthetic generators as the main evaluation and is not a substitute for ambient-trace replay.

**Drift and retries.** Figure 13(b) makes the guard tradeoff explicit. At $\rho$ = 20ppm, 5, 10, and 20 ms guards remain valid for 125, 250, and 500 s; at 100 ppm those ages fall to 25, 50, and 100 s. Table X prices $R_{\max}$ without assuming independent cycles: Eq. (15) bounds a needless fallback by $(1 - p_0)^{R_{\max}}$, while a true disconnection is recognized after at most $R_{\max} T_{cyc}$. The selected $R_{\max}$ = 3 caps that bound at 0.8% and the detection delay at three cycles.

**Capacitance and role density.** At fixed current and thresholds, ideal charge time scales with $C$: 100, 220, and 1000 $\mu$F take 2.13, 4.68, and 21.3 times the 47 $\mu$F reference interval. Below Eq. (4), an operation is infeasible; above it, storage trades fewer gate failures for a longer cold start. Role density is separate: Figure 10(b) empirically places the best tested point near $K = m$, consistent with Eq. (8). The fixed $K = 32$ firmware setting should therefore not be extrapolated to 120 contenders.

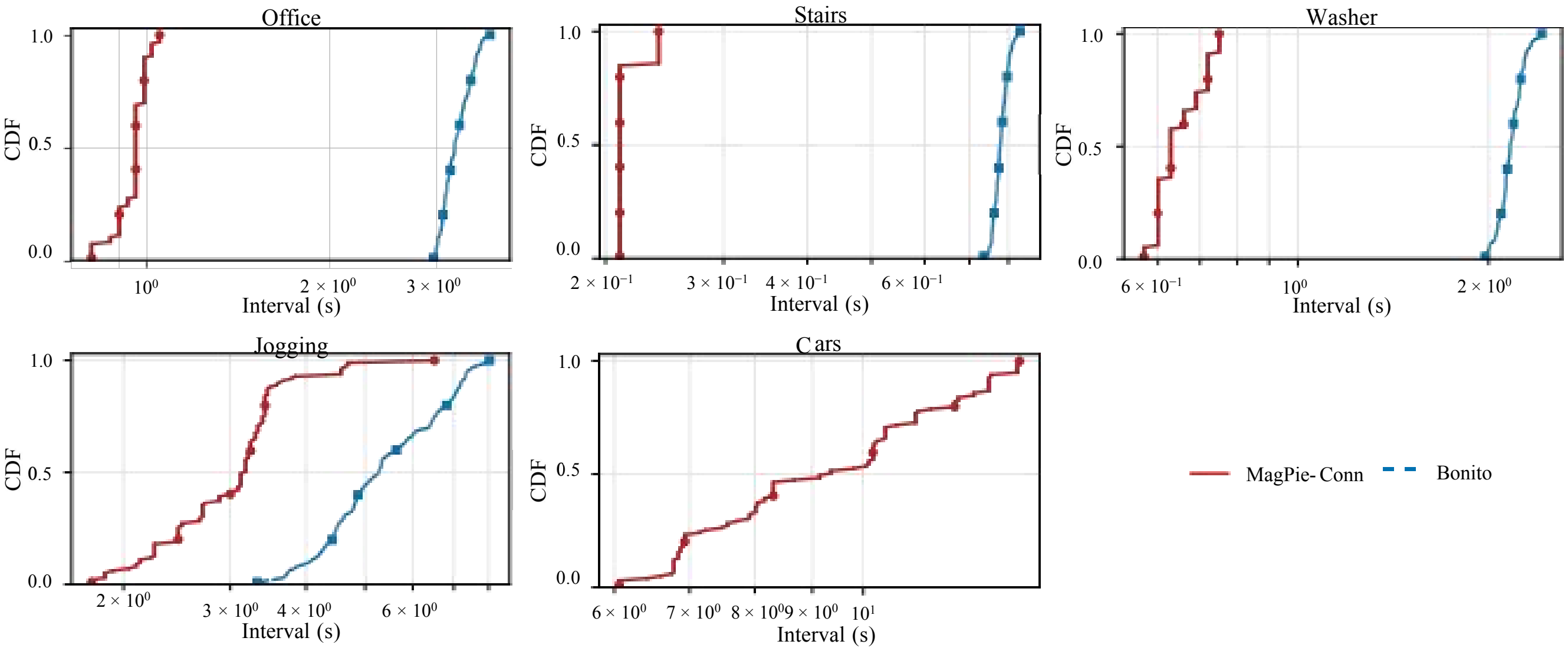

Fig. 9. CDF of sustained inter-exchange intervals on logarithmic horizontal axes after initial discovery. Bonito does not establish a Cars connection within 10,000 s.

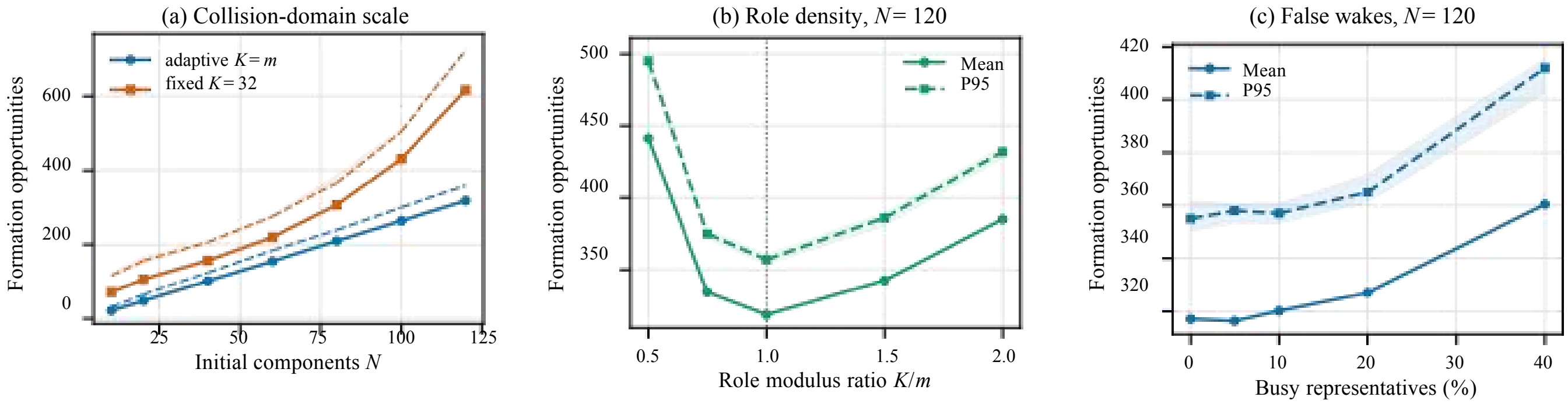

Fig. 10. MAC-only component formation over 500 trials. Solid lines are means; dashed lines are P95. Shading gives bootstrap 95% intervals for P95. (a) Adaptive $K = m$ versus the prototype's fixed $K = 32$; (b) role-density sweep; (c) a false wake makes one representative busy for one opportunity. All representatives are otherwise ready, and the model has one collision domain with no capture.

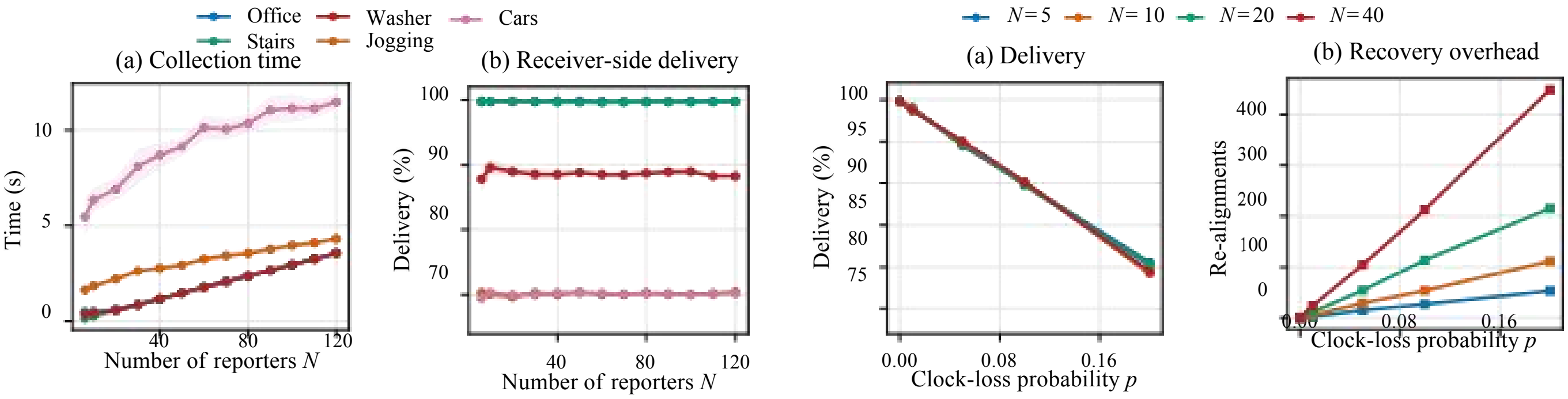

Fig. 11. Network energy-model scalability over ten seeds of 50 rounds. Lines are seed means; shading is the 95% confidence interval. Discovery uses matching and is excluded.

Fig. 12. Effect of per-cycle LP-RTC failure probability $p$ in Office (ten seeds). At $p = 0.2$, delivery is 79–81%, while re-alignments grow approximately linearly with $N$.

### F. Testbed Validation

**Pairwise synchronization.** With controlled charging times between 500 and 5000 ms, the logic traces contain 459 complete MagPie latency pulses in 3459 s and 40 Find pulses in 3315 s. Normalized rates are 478 and 43 completed alignments per hour, an 11.0× increase. Mean synchronization latency is 3678 versus 24,012 ms (6.5 × lower), and maxima are 34,405 versus 222,942 ms. Counts include only closed high pulses; rising and falling transitions are not treated as separate events. The smaller hardware gain relative to simulation is expected because the RF-powered testbed has a narrower energy range and includes real radio and firmware overhead.

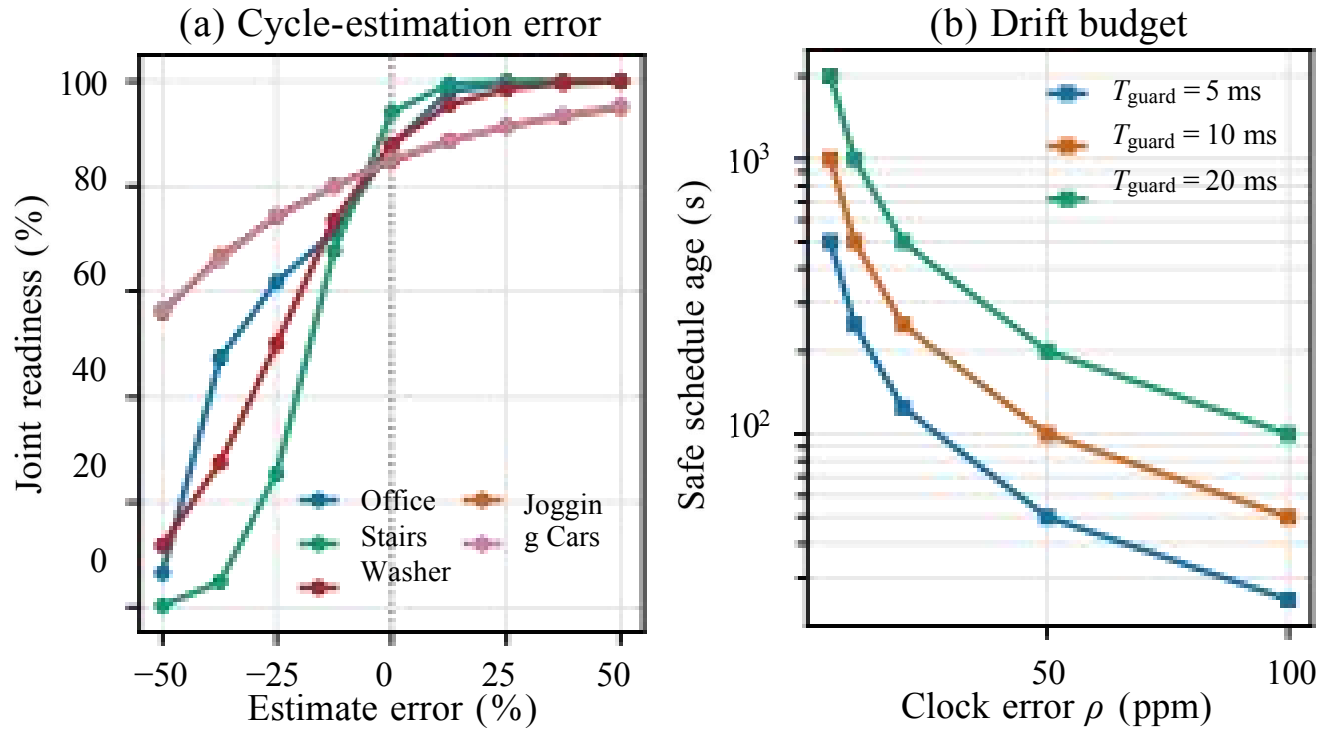

Fig. 13. Parameter sensitivity. (a) A multiplier perturbs the rolling 0.9-quantile cycle estimate; each point uses 4968 next-cycle decisions per scenario. (b) Safe schedule age $\tau_{stale} = T_{guard}/(2\rho)$ for three guard intervals.

TABLE X
RETRY SENSITIVITY AT $\alpha = 0.9$ ($p_0 = 0.8$).

| $R_{max}$ | 1 | 2 | **3** | 4 | 5 |
|---|---|---|---|---|---|
| Fallback bound | 20% | 4% | **0.8%** | 0.16% | 0.032% |
| Detection delay | $T_{cyc}$ | $2T_{cyc}$ | $3T_{cyc}$ | $4T_{cyc}$ | $5T_{cyc}$ |

**Network operation.** Five reporting boards and one sink run for 11.05 hours. Formation completes in 945 s, after which the sink logs 2998 received data packets with 13.2 s mean source-to-sink latency. The firmware log does not contain the denominator of scheduled transmission opportunities, so these records cannot yield a packet-delivery ratio. The controlled RF source and 1000 $\mu$F buffer validate execution of alignment, persistence, and slotting on real silicon; they do not validate ambient harvesting, large scale, or end-to-end energy.

## V. DISCUSSION AND LIMITATIONS

**Three parameters govern different risks.** The listening variables ($T_{listen}$, $\delta$) determine whether a beacon is observable, $K$ controls role collisions, and $\alpha$ trades scheduled delay against energy readiness. Equations (5), (8), and (12) expose these as separate controls rather than one tuning knob. Their guarantees remain conditional: a persistent beat frequency can defeat the uniform-phase approximation, and a regime shift can invalidate quantile coverage. The current firmware fixes the parameters; online joint adaptation is a design direction, not an evaluated result.

**Persistence is finite.** A backed LP-RTC changes the cost of a miss only while its domain survives and its drift stays within the guard. The leakage examples following Eq. (2) reduce a nominal 3.2-day ideal to hours, and the 100-ppm sensitivity shortens a 10-ms guard to 50 s. The forced-failure experiment shows the protocol consequence, but IID loss is not a physical failure model. Temperature-correlated drift or a shared reservoir fault could invalidate multiple edges at once and produce a more severe recovery burst.

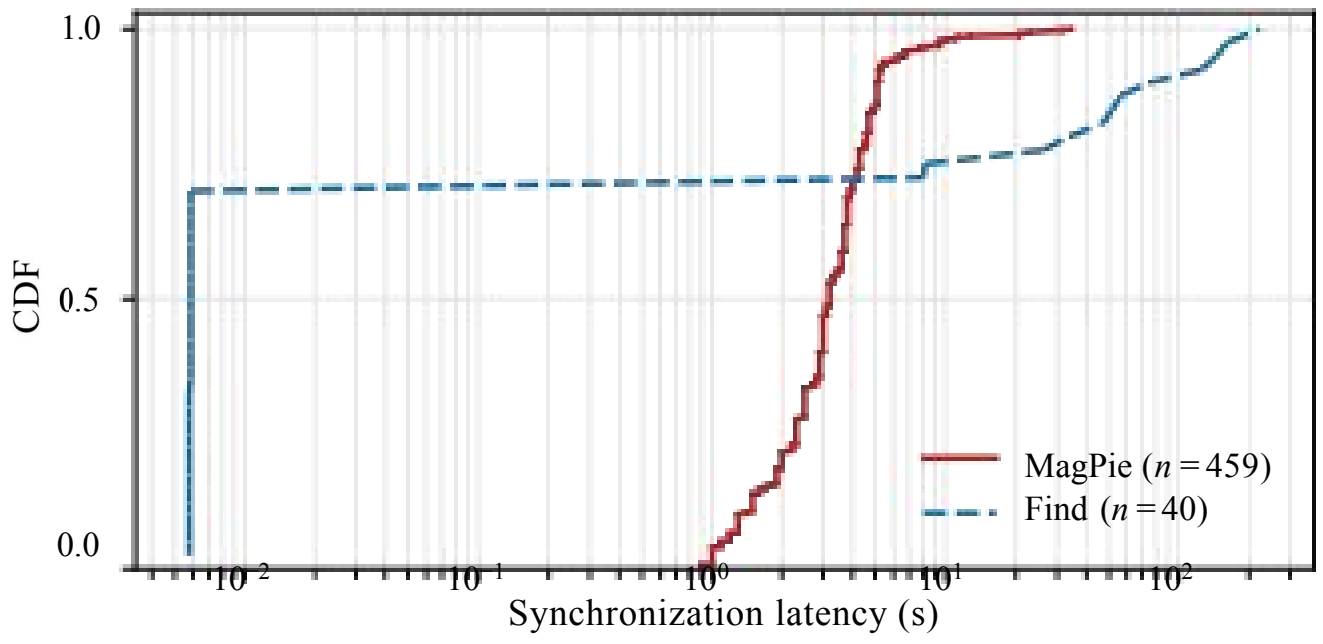

Fig. 14. Synchronization latency from logic-analyzer high-pulse widths: 459 completed MagPie events over 3459 s and 40 Find events over 3315 s.

**Energy accounting bounds the claim.** The 1000 $\mu$F testbed buffer provides implementation margin but takes 21.3 × the ideal charge time of the simulator's 47 $\mu$F reference over the same voltage interval. Equal simulated capacitance prevents that buffer from favoring MagPie, but it does not supply the missing shunt measurements. Converter efficiency, threshold circuitry, backup leakage, false wakes, and NVM commits can shift both thresholds and rankings. The present results therefore establish timing behavior and modeled primitive costs, not measured end-to-end energy efficiency.

**The powered reference is a system comparison.** The per-scenario coordinator oracle is faster in Stairs and slower in the other four parameterized settings. It simultaneously changes control infrastructure and removes the WuR, so it cannot attribute causality to either factor. A powered coordinator remains attractive where deployment and maintenance are acceptable. Completing the full WuR–clock–coordinator ablation grid on iso-storage hardware is necessary before claiming which primitive supplies each gain.

**Collection is limited by energy, not only slots.** Fixed slots arrive while a reporter may lack radio-grade energy, yielding approximately 70% receiver-side delivery in the two most variable settings even under an error-free channel. An energy-aware scheduler could order nodes by recent readiness, let the sink advertise unused slots, or reclaim a missed range in a later epoch. Such changes consume control energy and can break idempotence if a brownout hides a reassignment; they must therefore be versioned and evaluated rather than assumed to recover the missing 30%. One-way senders also lack delivery confirmation unless a cumulative ACK is added.

**Evidence has deliberate boundaries.** Controlled logic traces and the 11.05-hour six-device run validate execution, but the latter lacks an opportunity denominator and hence a delivery ratio. The Powercast source is energy infrastructure, not a scheduling anchor. Synthetic traces reproduce selected Bonito statistics but are not direct HDF5 replay. The network energy model assumes an available sink and error-free assigned slots; the separate contention study adds collisions and false busy periods, but assumes continuously eligible representatives, one collision domain, and no capture. None of these artifacts alone demonstrates field-scale RF operation.

**Multi-hop is not a mechanical extension.** In multiple collision domains, hidden terminals make one global $K$ insufficient, and capture or fading changes the solo-transmission model.

Each route would need persistent per-hop offsets and versioned forwarding epochs; delivery probability and guard energy would compound along the path. A parent clock loss can stall an entire subtree, while simultaneous low-energy parents can create cascading brownouts near the sink. The existing membership epoch and merge tree provide a recovery namespace, but not channel allocation, energy-aware parent selection, or loop-free repair under concurrent failures. Mobility further changes neighbors, link quality, and energy statistics together. These problems require a topology-aware MAC and a physical multi-hop deployment and are outside the claims of this paper.

## VI. Conclusion

MagPie demonstrates a specific co-design point: use a WuR to enlarge the first-contact opportunity and preserve the resulting phase in an independently backed clock. Its [illegible] [illegible] [illegible] depends on adapting $K$ to contender density. Controlled hardware validates protocol execution, but the present evidence does not establish end-to-end energy efficiency, field-scale RF behavior, or multi-hop operation. Those claims require shunt-based power measurements, direct ambient-trace replay, a complete reliability denominator, and larger deployments with interference and intermittent sinks.